\documentclass[11pt,a4paper,notitlepage]{article}
\usepackage{amssymb,amsmath,amstext,amsfonts,empheq,mdframed,xcolor}
\usepackage{bbold}
\usepackage{cite}
\usepackage{bm}
\usepackage{graphicx}
\usepackage{overpic}
\usepackage{subcaption}
\usepackage{braket}
\usepackage{hyperref}
\usepackage{textcomp}
\usepackage[margin=2.1cm]{geometry}

\usepackage{fetamont}
\usepackage[T1]{fontenc}

\definecolor{lgray}{gray}{0.90}
\newcommand*\graybox[1]{ \colorbox{lgray}{\hspace{1em}#1\hspace{1em}}} 

\newcommand{\Pzeta}{\mathcal{P}_\zeta}
\newcommand{\OGW}{\Omega_\textsc{gw}}
\newcommand{\kref}{k_\textrm{ref}}
\newcommand{\krh}{k_\textrm{rh}}
\newcommand{\TRD}{\mathcal{T}_\textrm{rad}}
\newcommand{\Tw}{\mathcal{T}_w}
\begin{document}
\title{{\Huge {\fontfamily{ffm}\selectfont \href{https://github.com/Lukas-T-W/SIGWfast}{SIGWfast}}} \\ A python package for the computation of \\
scalar-induced gravitational wave spectra}
\author{Lukas T.~Witkowski \\ \href{https://github.com/Lukas-T-W/SIGWfast}{https://github.com/Lukas-T-W/SIGWfast} \\
{\small \href{mailto:lukas.witkowski@iap.fr}{lukas.witkowski@iap.fr}}}
\date{}

\maketitle

\begin{center}
\textit{Institut d'Astrophysique de Paris, GReCO, UMR 7095 du CNRS et de Sorbonne Universit\'{e},\\ 98bis boulevard Arago, 75014 Paris, France}

\vspace{1.5cm}

\end{center}

\begin{abstract}
\noindent \textbf{SIGWfast} is a python code to compute the \textbf{S}calar-\textbf{I}nduced \textbf{G}ravitational \textbf{W}ave spectrum from a primordial scalar power spectrum that can be given in analytical or numerical form. SIGWfast was written with the aim of being easy to install and use, and to produce results fast, typically in a matter of a few seconds. To this end the code employs vectorization techniques within python, but there is also the option to compile a C++ module to perform the relevant integrations, further accelerating the computation. The python-only version should run on all platforms that support python 3. The version employing the C++ module is only available for Linux and MacOS systems.
\end{abstract}

\vspace{0.5cm}

\begin{figure}[h]
\centering
\includegraphics[width=0.6\textwidth,angle=0]{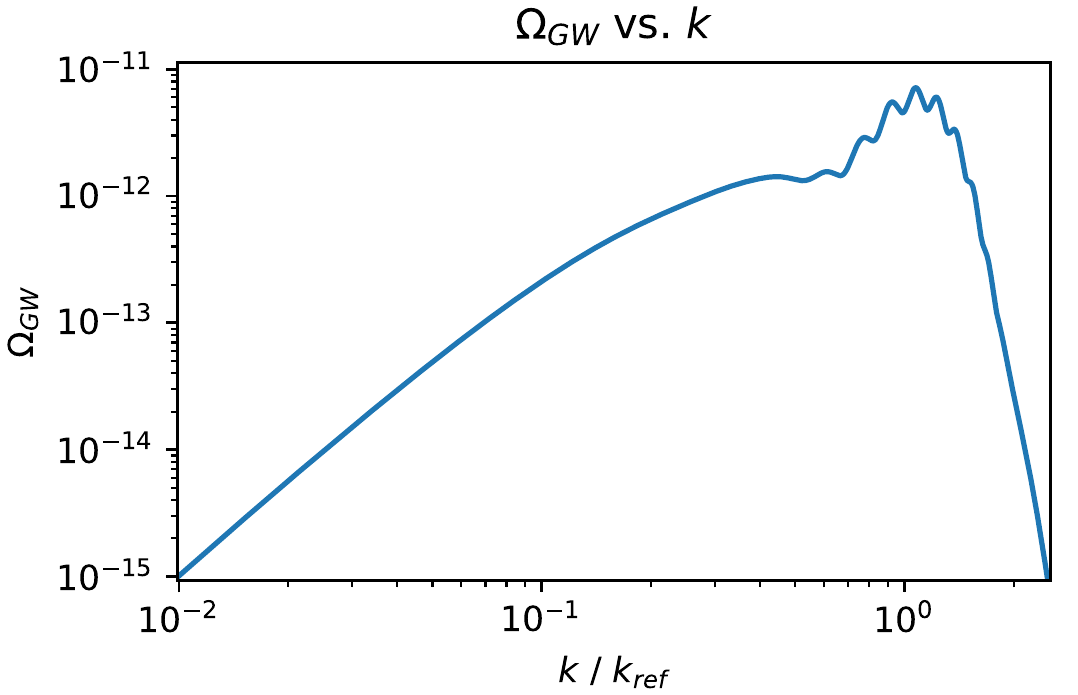}
\end{figure}

\newpage

\tableofcontents

\section{Introduction}
The nascent era of gravitational wave astronomy offers new ways of testing cosmic inflation through its signature in the stochastic gravitational wave background \cite{Achucarro:2022qrl,LISACosmologyWorkingGroup:2022jok}. An important observable in this context is the spectrum of scalar-induced gravitational waves (SIGW), see \cite{Domenech:2021ztg} for a recent review, which is produced when scalar fluctuations that exited the horizon during inflation re-enter in the post-inflationary era.\footnote{Scalar fluctuations also induce gravitational waves during inflation, see e.g.~\cite{Fumagalli:2021mpc}, which can be promising targets for future detection efforts. Here, SIGW refers exclusively to gravitational waves induced during the post-inflationary era, which is what SIGWfast computes.}

For the SIGW signal to be potentially detectable by the upcoming generation of gravitational wave observatories, the sourcing scalar fluctuations need to be sufficiently large. At large scales ($\gtrsim 1 \textrm{ Mpc}$) the amplitude of scalar fluctuations is bounded by Cosmic Microwave Background (CMB) and Large Scale Structure (LSS) data, and the corresponding SIGW signal will be undetectable. However, these bounds do not apply at small scales ($\ll 1 \textrm{ Mpc}$), allowing for a significant enhancement of scalar fluctuations, rendering the SIGW signal potentially observable. As a result, the SIGW spectrum is particularly suited for testing inflation at small scales, thus complementing CMB and LSS measurements, which give information 
about large scales. 

Inflation models that enhance scalar fluctuations have been studied extensively in recent years. One reason is the increased interest in production mechanisms for primordial black holes (see \cite{Sasaki:2018dmp, Garcia-Bellido:2018leu, Carr:2020gox} for overviews of the topic), which are generated as overdensities sourced by large scalar fluctuations collapse. In addition, such models have also been studied in their own right as completions of inflation at small scales, where constraints from CMB and LSS data are not available, and inflation can depart from the single-field slow-roll paradigm favoured at large scales. In all these models the spectrum of SIGW is an important observable, whose detection could give important information about the mechanism of inflation.\footnote{To give an example, a characteristic signature of a departure of inflation from the single-field slow-roll paradigm is an oscillatory modulation of the SIGW spectrum. Its properties can be directly related to inflationary-era quantities \cite{Fumagalli:2020nvq, Fumagalli:2021cel}.}

As a result, the computation of the spectrum $\OGW(k)$ of the SIGW is in the process of becoming a routine task for inflationary model builders with an interest in mechanisms beyond single-field slow-roll models. The leading contribution to the SIGW spectrum is computed as a convolution-like double integral over two factors of the scalar power spectrum $\Pzeta(k)$ and a transfer function $\mathcal{T}$, i.e.~$\OGW \sim \iint \mathcal{T} \Pzeta \Pzeta$ \cite{Ananda:2006af,Baumann:2007zm,Espinosa:2018eve,Kohri:2018awv}. Even though the transfer functions are known explicitly for various post-inflationary histories \cite{Espinosa:2018eve,Kohri:2018awv,Domenech:2019quo,Domenech:2020kqm}, the integration typically has to be performed numerically, and analytic results can only be derived for idealised choices of $\Pzeta$ (like $\delta$-distributions \cite{Kohri:2018awv,Cai:2019amo} or lognormal peaks \cite{Pi:2020otn}), or in certain limits (e.g.~in the IR limit $k \rightarrow 0$). Thus every researcher on this topic is eventually confronted with the task of computing this integration numerically. While performing a double integral numerically is not a priori difficult, there are a couple of subtleties to be considered in this case. Firstly, the transfer function $\mathcal{T}$ typically exhibits a divergence, and the dominant contribution to the integral comes from integrating over the vicinity of the divergent locus. Hence all-purpose numerical integration routines like \texttt{NIntegrate} of Wolfram Mathematica or \texttt{scipy.integrate.quad} in python have to be used with care to treat the singular locus correctly. In the worst case, the computation produces an error or, even worse, wrong results. Secondly, one of the integrations is unbounded above, so that an upper limit has to be set in practice. This has to be chosen correctly to not affect the result. Finally, as the double integral has to be evaluated for every value of $k$, an unoptimized code can run significantly longer (by a factor of 100 or more!) than a version that has been designed for high speed.

These considerations are strong motivations for a public code to compute the SIGW spectrum that is \emph{(i)} easy to use, \emph{(ii)} robust against possible pitfalls, and \emph{(iii)} computes $\OGW(k)$ fast. The package `SIGWfast' has been written with these three goals in mind. To be easy to use, SIGWfast has a python interface where the computation can be configured and through which the input scalar power spectrum $\Pzeta(k)$ is provided. It can also be run on all systems that have a functioning installation of python 3. For robustness, SIGWfast sets the integration cutoff automatically. It also makes very little demands on how the scalar power spectrum is to be provided.\footnote{For example, if $\Pzeta$ is to be given by an analytic formula, this does not have to be a vectorizable function.} It also outputs plots which allow the user to check that the computation proceeded correctly. Regarding speed, SIGWfast comes as a python-only version, which uses vectorization and list comprehension to minimise computation time. It also provides the option to use a compiled C++ module for performing the integration, which further reduces the computation time by up to 25\%. This option is however restricted to systems running on Linux or MacOS. Overall, on a modern laptop or desktop machine a computation of $\OGW(k)$ with a quality suitable for publication typically takes $\mathcal{O}(1)$ seconds for the python-only version as well as the version using the compiled C++ module.

SIGWfast can be downloaded from \href{https://github.com/Lukas-T-W/SIGWfast/releases}{here}. It is free software that is distributed under the MIT License: This means that you can use, copy, modify, merge, publish and distribute, sublicense, and/or sell copies of it, and to permit persons to whom it is furnished to do so it under the terms of the MIT License. If you use SIGWfast, we kindly ask you to cite this arXiv article in any resulting publications. 

The rest of this document contains a user guide that is structured as follows. In sec.~\ref{sec:output} we explain in detail what SIGWfast computes. Sec.~\ref{sec:prereq-inst-config} is dedicated to instructions for how to install, configure and use SIGWfast. In sec.~\ref{sec:examples} we show a couple of example computations. Some potential issues and their solution are listed in sec.~\ref{sec:trouble}. A link to the MIT License can be found in sec.~\ref{sec:license}. We hope that you will find SIGWfast useful!

\section{What does SIGWfast compute?}
\label{sec:output}
The code computes the energy density fraction $\OGW(k)$ in gravitational waves today that was induced by scalar fluctuations with primordial power spectrum $\Pzeta(k)$. As this depends on the equation of state of the universe when the gravitational waves are produced, the package SIGWfast contains two python scripts, \texttt{SIGWfast.py} and \texttt{SIGWfastEOS.py}, to allow for exploring different expansion histories of the universe after inflation:
\begin{itemize} 
\item \texttt{SIGWfast.py} computes the gravitational wave spectrum induced during a period of radiation domination, which we expect to be the result of interest for most users. 
\item \texttt{SIGWfastEOS.py} computes the gravitational wave spectrum induced during a period of the universe with equation of state parameter $w$, which can be set by the user. As the integration kernel is more involved to allow for different values of $w$, this script is slightly slower than \texttt{SIGWfast.py}.
\end{itemize}
In the following, we describe in detail the output produced by the two scripts.

\subsection{\texttt{SIGWfast.py}: gravitational waves induced during radiation domination}
\label{sec:output-SIGWfast}
For gravitational waves induced during a period of radiation domination, the leading contribution to the fraction of energy density in gravitational waves is given by \cite{Ananda:2006af, Baumann:2007zm}:\footnote{\label{ftn:trispectrum}Strictly speaking, the leading scalar-induced contribution to $\OGW$ depends on the four-point function of scalar fluctuations $\langle \zeta \zeta \zeta \zeta \rangle$. This can then be split into two pieces, one involving $\Pzeta \Pzeta$, which is what is shown in \eqref{eq:OmegaGW-rad}, and an additional piece involving the primordial trispectrum \cite{Unal:2018yaa}, which we ignore here. The reason is that for most well-behaved inflation models (i.e.~models that remain within the bounds of perturbative control) the piece with $\Pzeta \Pzeta$ dominates so that the contribution from the trispectrum can be safely ignored \cite{Garcia-Saenz:2022tzu}.}
\begin{align}
\label{eq:OmegaGW-rad}
    \OGW^{\textrm{rad}}(k) = \mathcal{N} \int_0^1 \textrm{d} d \int_1^\infty \textrm{d} s \, \TRD(d,s) \, \Pzeta \bigg(\frac{k}{2}(s+d)\bigg) \Pzeta \bigg(\frac{k}{2}(s-d)\bigg) \, .
\end{align}
with \cite{Espinosa:2018eve,Kohri:2018awv}
\begin{align}
\label{eq:TRDds}
\TRD(d,s) = 12 \frac{{\big(d^2-1\big)}^2{\big(s^2-1\big)}^2{\big(d^2+s^2-6\big)}^4}{{\big(s^2-d^2\big)}^8} \bigg[ \bigg(\ln \frac{3-d^2}{|s^2-3|} + \frac{2 \big(s^2-d^2\big)}{d^2+s^2-6}\bigg)^2 + \pi^2 \Theta\big(s-\sqrt{3}\big) \bigg] .
\end{align}
To get the spectrum of gravitational waves today, the prefactor $\mathcal{N}$ needs to be set to $\mathcal{N}= c_g \Omega_{\textrm{r},0}$, where $\Omega_{\textrm{r},0} = 8 \cdot 10^{-5}$ corresponds to the energy density fraction in radiation today and $c_g$ depends on the number of relativistic species during the time of gravitational waves generation vs.~today as 
\begin{align}
c_g = \frac{g_{*,\textrm{rad}}}{g_{*,0}} \bigg( \frac{g_{*,0}}{g_{*S,\textrm{rad}}} \bigg)^{4/3} \, .
\end{align}
If the matter content of the universe is that of the Standard Model only, and during gravitational wave production all particles were relativistic, one finds $c_g \approx 0.4$. To allow for different values, in \texttt{SIGWfast.py} the prefactor  $\mathcal{N}$ is implemented as the variable \texttt{norm} that can be set by the user.\footnote{That is, setting \texttt{norm = 1} in \texttt{SIGWfast.py} corresponds to setting $\mathcal{N}=1$ in \eqref{eq:OmegaGW-rad}.} 

The principal input for \texttt{SIGWfast.py} is the primordial power spectrum $\Pzeta(k)$ that can be provided in numerical or analytical form. The script then computes the result of expression \eqref{eq:OmegaGW-rad}, i.e.
\begin{empheq}[box=\graybox]{align}
    \textrm{Output } \OGW^{\textrm{out}} \textrm{ of \texttt{SIGWfast.py}:} \quad \OGW^{\textrm{out}}(k) = \OGW^{\textrm{rad}} (k) \, .
\end{empheq}
The result is saved in a \texttt{.npz} file in the `data' subdirectory containing the user-specified array of $k$-values and the corresponding values of $\OGW^{\textrm{out}}$. In addition, both $\Pzeta(k)$ and $\OGW(k)$ are plotted. 

\subsection{\texttt{SIGWfastEOS.py}: gravitational waves induced during a phase with equation of state $w$}
\label{sec:output-SIGWfastEOS}
If inflation is not immediately followed by the hot big bang phase, scalar-induced gravitational waves can also be produced during an era where the universe is not dominated by radiation. One possibility is to consider a universe that before the epoch of radiation domination is described by a constant equation of state, parameterised by $w$. For gravitational waves induced during such a period, the contribution to the energy density fraction in gravitational waves is given by \cite{Domenech:2019quo,Domenech:2020kqm,Domenech:2021ztg}:\footnote{Again, we ignore the contribution from the primordial trispectrum for reasons as explained in footnote \ref{ftn:trispectrum}.} 
\begin{align}\label{eq:OmegaGW-w} 
\OGW^w(k)&= \mathcal{N} \, {\left(\frac{k}{\krh}\right)^{-2b}} \int_0^1 \textrm{d} d \int_1^\infty \textrm{d} s \, \Tw(d,s) \, \Pzeta \bigg(\frac{k}{2}(s+d)\bigg) \Pzeta \bigg(\frac{k}{2}(s-d)\bigg) \, ,
\end{align}
with
\begin{align}
b \equiv \frac{1-3w}{1+3w} \, .
\end{align}
The integration kernel $\Tw(d,s)$ takes the form \cite{Domenech:2019quo,Domenech:2020kqm,Domenech:2021ztg}:
\begin{align}\label{eq:Twcs}    
\Tw(d,s) = & \hphantom{\times} \mathcal{F}(b) {\bigg(\frac{(d^2-1)(s^2-1)}{d^2-s^2}\bigg)}^2 \, \frac{|1-y^2|^b }{(s^2-d^2)^2} \, \times \\
\nonumber & \times \Bigg\{\hphantom{+} \hphantom{\frac{4}{\pi^2}} \bigg[\mathsf{P}_{b}^{-b}(y) + \frac{2+b}{1+b} \mathsf{P}_{b+2}^{-b}(y) \bigg]^2 \Theta \big(s-c_s^{-1} \big) \\
\nonumber &\hphantom{\times \Bigg\{} +  \frac{4}{\pi^2} \bigg[\mathsf{Q}_{b}^{-b}(y) + \frac{2+b}{1+b} \mathsf{Q}_{b+2}^{-b}(y) \bigg]^2 \Theta \big(s-c_s^{-1} \big) \\
\nonumber &\hphantom{\times \Bigg\{} +  \frac{4}{\pi^2} \bigg[\mathcal{Q}_{b}^{-b}(-y) + 2 \frac{2+b}{1+b} \mathcal{Q}_{b+2}^{-b}(-y) \bigg]^2 \Theta \big(c_s^{-1}-s \big) \Bigg\} \, ,
\end{align}
with
\begin{align}
    \label{eq:ydef}
y \equiv \frac{s^2+d^2-{2}{c_s^{-2}}}{s^2-d^2} \, , \quad 
    \mathcal{F}(b)=\frac{1}{3}\left( \frac{4^{1+b}(b+2)}{\left(1+b\right)^{1+b}(2b+3)c_s^2} \, \Gamma^2 \Big[b+\tfrac{3}{2} \Big] \right)^2\, ,
\end{align}
where $c_s$ is the propagation speed of scalar fluctuations, $\mathsf{P}^{\mu}_{\nu}(x)$ and $\mathsf{Q}^{\mu}_{\nu}(x)$ are the Ferrers function of the first and second kind, respectively, and $\mathcal{Q}^{\mu}_{\nu}(x)$ is the associated Legendre function of the second kind. We recorded their definitions in appendix \ref{app:Legendre-definition}.

The factor $(k/\krh)^{-2b}$ accounts for the fact that before the transition to radiation domination, gravitational waves redshift differently from the bulk. The `reheating scale' $\krh$ corresponds to the comoving wavenumber that entered the horizon at the transition to radiation domination, i.e.~$\krh=a_{\rm rh}H_{\rm rh}$, and is in general different for different cosmological models. What is important is that \eqref{eq:OmegaGW-w} is valid for $k \gg \krh$, i.e.~for modes that entered the horizon well before reheating, see e.g.~\cite{Domenech:2021ztg} for details. That is, we should not use \eqref{eq:OmegaGW-w} and hence \texttt{SIGWfastEOS.py} to compute the SIGW with $k \lesssim \krh$. As in \texttt{SIGWfast.py}, the normalisation factor $\mathcal{N}$ is implemented as the variable \texttt{norm} whose value can be specified by the user. To obtain the energy density fraction in gravitational waves today, this should be set to $\mathcal{N}= c_g \Omega_{\textrm{r},0}$, see the discussion in sec.~\ref{sec:output-SIGWfast} for more details. For $\mathcal{N}=1$ the expression in \eqref{eq:OmegaGW-w} corresponds to the gravitational wave spectrum right after the transition to radiation domination.

The principal inputs for \texttt{SIGWfastEOS.py} are the value of $w$ and the primordial power spectrum $\Pzeta(k)$, which can be provided in numerical or analytical form. The code \texttt{SIGWfastEOS.py} then computes the gravitational spectrum induced for either a universe described by an adiabatic perfect fluid ($c_s^2=w$) or a universe dominated by a canonical scalar field ($c_s^2=1$). The user can choose between these two options by setting the flag \texttt{cs\_equal\_one} to \texttt{False} or \texttt{True}, respectively. In the adiabatic perfect fluid case the value of $w$ needs to be chosen in the interval $0<w<1$ for the kernel in \eqref{eq:Twcs} to be valid, and \texttt{SIGWfastEOS.py} only permits values in this range. To reproduce the results of \texttt{SIGWfast.py} for radiation domination, in \texttt{SIGWfastEOS.py} one would need to set $c_s^2=w=1/3$ (i.e.~\texttt{w = 1/3} and \texttt{cs\_equal\_one = False}).\footnote{For $c_s^2=w=1/3$ the kernel $\Tw$ in \eqref{eq:Twcs} exhibits divergent pieces, which one can show to cancel analytically, so that the kernel reduces to $\TRD$ in \eqref{eq:TRDds} as expected. To deal with these unphysical divergences numerically, when inputting a value $w$ with $|w-1/3| < \epsilon$ in \texttt{SIGWfastEOS.py}, the value of $w$ is redefined automatically to $w=1/3-\epsilon$ with $\epsilon=10^{-4}$. This avoids the divergences without changing the result in 
a significant way.} Instead of providing $\krh$ as a further input, the script \texttt{SIGWfastEOS.py} computes \eqref{eq:OmegaGW-w}, but with the $\krh$-dependent term scaled out, i.e.
\begin{empheq}[box=\graybox]{align}
    \textrm{Output } \OGW^{\textrm{out}} \textrm{ of \texttt{SIGWfastEOS.py}:} \quad \OGW^{\textrm{out}}(k) = \big(\kref / \krh \big)^{2 b} \times \OGW^{w} (k) \, .
\end{empheq}
Here $\kref$ is the reference unit for wavenumbers $k$ as chosen by the user.\footnote{The scale $\kref$ is defined in the sense that writing \texttt{k = 0.1} in the python script corresponds to $k = 0.1 \kref$.} As the factor $\big(\kref / \krh \big)^{2 b}$ is a constant, its only effect is an overall rescaling which does not affect the spectral shape of $\OGW^{w}$ as a function of $k$. 

Again, the result is saved in a \texttt{.npz} file in the `data' subdirectory containing the user-specified array of $k$-values and the corresponding values $\OGW^{\textrm{out}}$. In addition, both $\Pzeta(k)$ and $\OGW(k)$ are plotted. 

\section{Prerequisites, installation and configuration}
\label{sec:prereq-inst-config}
\subsection{Prerequisites}
SIGWfast can be used on any system that supports python 3. We recommend using python environments and a package manager such as `conda'. The following python modules are required:\footnote{The modules `time' and `tqdm' are for timing the computation and displaying a progress bar, respectively, and could be dispensed with by commenting out appropriate lines of code.}
\begin{empheq}[box=\graybox]{align}
\nonumber
\textrm{math, matplotlib, numpy, os, scipy, sys, time, tqdm.} 
\end{empheq}

\vspace{0.3cm}

\noindent \textbf{Optional C++ extension:} This is only supported on systems running on Linux or MacOS. Compiling the C++ extension further requires the modules:
\begin{empheq}[box=\graybox]{align}
\nonumber
\textrm{distutils, platform, shutil, subprocess,}  
\end{empheq}
and a working C++ compiler. 

\subsection{Installation}
To use SIGWfast, download the latest release as a \texttt{.zip} or \texttt{.tar.gz} archive from \href{https://github.com/Lukas-T-W/SIGWfast/releases}{this page}. After decompression the necessary files and directory structure are already in place in the parent directory.

\subsection{File content}
The parent directory contains two python scripts \texttt{SIGWfast.py} and \texttt{SIGWfastEOS.py}, whose execution computes the result. Their respective outputs are described in detail in sec.~\ref{sec:output}.

The subdirectory `libraries' contains files necessary for performing the computation and which do not need to be modified by the user:
\begin{itemize}
\item \texttt{sdintegral.py} contains the definitions and kernels for computing the relevant integrals.
\item \texttt{SIGWfast.cpp} is a C++ script to perform the integrals in the computation of $\OGW$.
\item \texttt{setup.py} is the code that configures and executes the compilation of the python module that will execute the C++ code contained in \texttt{SIGWfast.cpp}. The resulting python module named `sigwfast' is also deposited in the `libraries' subdirectory. The script \texttt{setup.py} is only called if the option to use the C++ extension is chosen by the user. 
\end{itemize}

The subdirectory `data' will receive the result data for $\OGW(k)$ in a \texttt{.npz} file. Also, if the input is a scalar power spectrum in numerical form, this needs to be provided in the `data' subdirectory as a \texttt{.npz} file. As an example, the file \texttt{P\_of\_k.npz} with data for $\Pzeta(k)$ for the inflation model \eqref{eq:std-example} is included.

\begin{figure}[t]
\centerline{
\includegraphics[width=0.62\textwidth,angle=0]{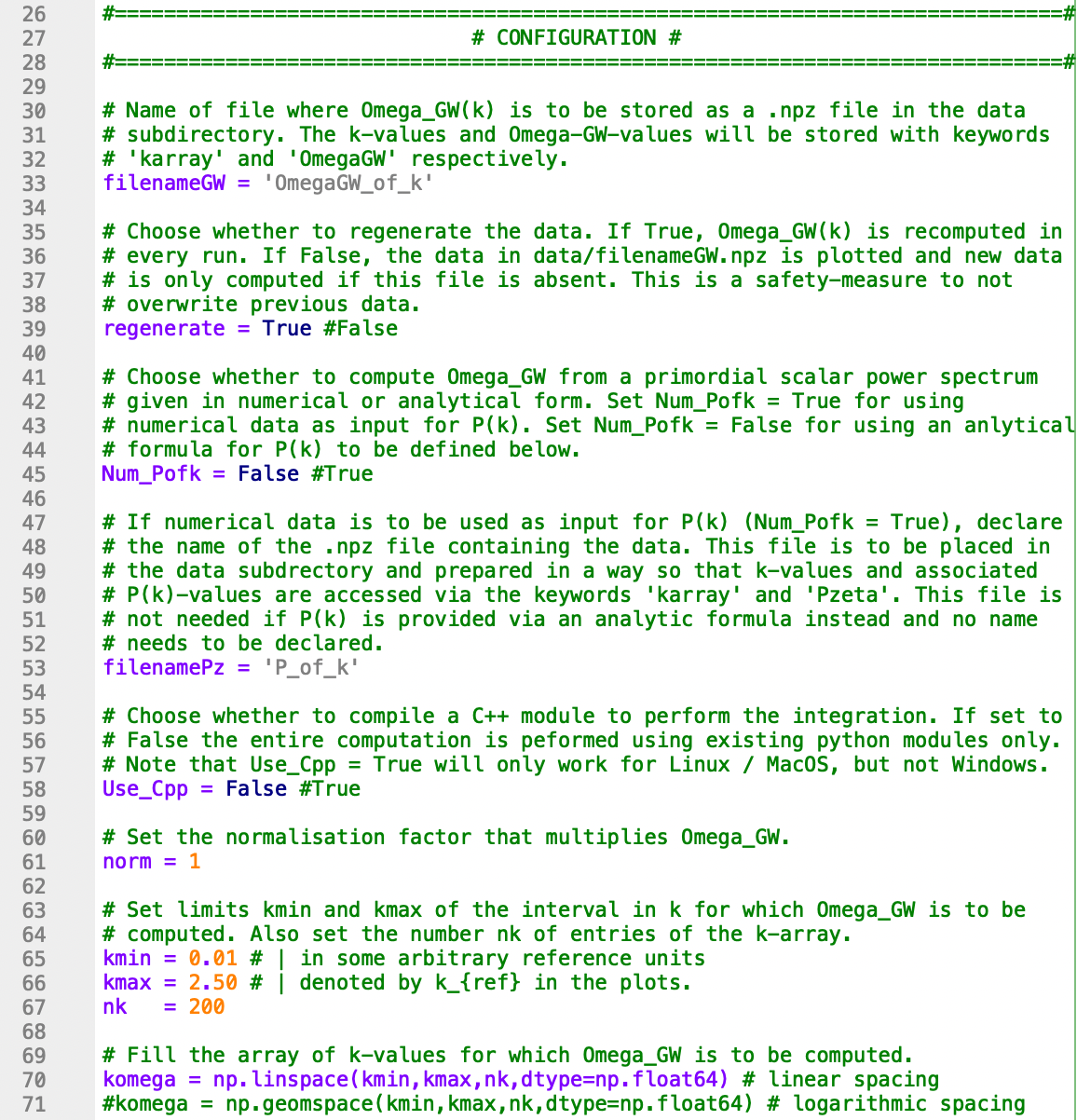}}
\caption{Configuration section of \texttt{SIGWfast.py} for the computation of the result in fig.~\ref{fig:P-and-OmegaGW-0p5-14-1}. To compute the same result using the compiled C++ module, set the flag \texttt{Use\_Cpp = True}.}
\label{fig:config-rad}
\end{figure}

\subsection{Quick guide}
Set flags and values for input parameters in the block of code titled `Configuration'. Provide the scalar power spectrum either by defining a function \texttt{Pofk(k)} in the block of code titled `Primordial scalar power spectrum' or in form of numerical data in a file \textquotesingle data/\textquotesingle +filenamePz+\textquotesingle .npz\textquotesingle. See the more detailed guide below for how this file is to be prepared. After this, you're good to go!

\subsection{Configuration of \texttt{SIGWfast.py} and \texttt{SIGWfastEOS.py}: step-by-step guide}
\label{sec:config-step-by-step}
This is a set of detailed instructions for configuring \texttt{SIGWfast.py} and \texttt{SIGWfastEOS.py}. The first seven steps are universal to both scripts. To this end open \texttt{SIGWfast.py} or \texttt{SIGWfastEOS.py} and go to the block of code labeled `Configuration', which is the first block after the header where the necessary modules are imported. In \texttt{SIGWfast.py} the relevant section of code looks like in fig.~\ref{fig:config-rad}. This is where you can adjust the script for your purposes. The necessary steps are as follows: 
\begin{enumerate}
\item Set \texttt{filenameGW}. Choose a name for the \texttt{.npz} file that will contain the results for $\OGW(k)$ and that will be deposited in the `data' subdirectory. The $k$-values and corresponding $\OGW$-values in this file will be accessible via the keywords `karray' and `OmegaGW'.\footnote{That is, to access the data in this file, load it via \texttt{data  = numpy.load(\textquotesingle data/\textquotesingle +filenameGW+\textquotesingle .npz\textquotesingle)}. The arrays with the values of $k$ and $\OGW$ are then given by \texttt{data[\textquotesingle karray\textquotesingle]} and \texttt{data[\textquotesingle OmegaGW\textquotesingle]}, respectively.} Note that if a file with this name already exists, a new run will in general overwrite the old file. To avoid this, see the next step.
\item Set the flag \texttt{regenerate}. If this is set to \texttt{True}, a run of the code will execute a new computation and save the result in the file \textquotesingle data/\textquotesingle +filenameGW+\textquotesingle .npz\textquotesingle, possibly overwriting an old file of the same name. If the flag is set to \texttt{False}, after hitting run, the code checks whether a file \textquotesingle data/\textquotesingle +filenameGW+\textquotesingle .npz\textquotesingle~already exists. If this is the case, no new computation is performed and instead the data in the existing file is plotted. This is a safety-measure to avoid existing data to be overwritten by accident. If however \textquotesingle data/\textquotesingle +filenameGW+\textquotesingle .npz\textquotesingle~does not exist, the code proceeds to performing the computation and saving the new data.
\item Set the flag \texttt{Num\_Pofk}. The code computes the scalar-induced gravitational wave spectrum using the primordial scalar power spectrum $\Pzeta(k)$ as input. $\Pzeta(k)$ can be provided in terms of numerical data or an analytic formula, the choice of which is declared by specifying the flag \texttt{Num\_Pofk}. If set to \texttt{True}, $\OGW$ will be computed from a scalar power spectrum given by the numerical data in a \texttt{.npz} file in the `data' subdirectory. The name of the file can be specified in the next step. If the flag \texttt{Num\_Pofk} is instead set to \texttt{False}, $\OGW$ will be computed from a scalar power spectrum that needs to declared explicitly as a function \texttt{Pofk(k)}. See sec.~\ref{sec:Pzeta} for for detailed instructions on how this is to be done.
\item Optional: declare \texttt{filenamePz}. In case $\OGW(k)$ is to be computed from numerical data (\texttt{Num\_Pofk = True}), give here the name of the \texttt{.npz} file located in the `data' subdirectory. The file is to be prepared so that $k$-values and associated $\Pzeta(k)$-values are to be accessed via the keywords `karray' and `Pzeta', respectively. If $\Pzeta(k)$ is to be provided via an analytic formula instead (\texttt{Num\_Pofk = False}), no input file is needed and this line of code is ignored.
\item Set the flag \texttt{Use\_Cpp}. If set to \texttt{True}, the code imports methods from the compiled module `sigwfast' to perform the integration, or, if the module does not yet exist, initiates a compilation of it from `libraries/SIGWfast.cpp'. This option is only available for Linux and MacOS systems and requires a functioning C++ compiler in addition to python. The module `sigwfast' only needs to be compiled once and can then be used for all further computations.\footnote{The reason is that `sigwfast' contains methods to perform in the integrations over $d$ and $s$ in \eqref{eq:OmegaGW-rad} and \eqref{eq:OmegaGW-w}, but is ignorant of the integrands.} That is, if the input power spectrum $\Pzeta(k)$ is changed (and/or in the case of \texttt{SIGWfastEOS.py} the parameter $w$) the originally compiled module can still be used and does \emph{not} need to be recompiled. If \texttt{Use\_Cpp} is instead set to \texttt{False}, the entire computation is done within python, using only the modules listed above under `Prerequisites'.
\item Set \texttt{norm}. This is the ``normalization'' factor $\mathcal{N}$ in \eqref{eq:OmegaGW-rad} and \eqref{eq:OmegaGW-w}. For $\mathcal{N}=1$ (i.e.~\texttt{norm = 1}) the output of \texttt{SIGWfast.py} and \texttt{SIGWfastEOS.py} corresponds to the energy density fraction in gravitational waves at the time of radiation domination. To get the corresponding result for today requires setting $\mathcal{N}=c_g \Omega_{\textrm{r},0}$, see sec.~\ref{sec:output} above. 
\item Declare the range of wavenumbers $k$ for which $\OGW(k)$ is to be computed. The corresponding array is called \texttt{komega} accordingly.  In SIGWfast, the default way of doing this is to define both a lower limit \texttt{kmin}, an upper limit \texttt{kmax} and setting the number of entries \texttt{nk} of \texttt{komega}. The values of $k$ can be specified in any unit of choice. The default option is then to fill \texttt{komega} with values that are linearly spaced (\texttt{numpy.linspace}), but logarithmic spacing (\texttt{numpy.geomspace}) or any other method are also allowed, with the only restriction that \texttt{komega} should be a numpy array. 

Note that for an analytic primordial power spectrum as input (\texttt{Num\_Pofk = False}) a good guideline is to choose \texttt{komega} such that $\Pzeta(k)$, when sampled over \texttt{komega}, exhibits all relevant features of the full scalar power spectrum.\footnote{This is also good practice from the physics point-of-view, as one expects $\OGW(k)$ and $\Pzeta(k)$ to exhibit structure on roughly the same scales.} The reason is that for better robustness and performance, $\Pzeta(k)$ is first discretized over an array \texttt{kpzeta}, before an interpolation is used in the computation. By default, in SIGWfast \texttt{kpzeta} is constructed from \texttt{komega} by extending its range and increasing the density of points, which should be sufficient for most applications. For specialised applications \texttt{kpzeta} can also be defined independently from \texttt{komega}, see sec.~\ref{sec:Pzeta} for details, so that \texttt{komega} can be chosen freely. 
\end{enumerate}
There are two further configuration steps needed for \texttt{SIGWfastEOS.py}:
\begin{enumerate}
\setcounter{enumi}{7}
\item Set the value of \texttt{w}. In \texttt{SIGWfastEOS.py} we also need to specify a value for the equation of state parameter $w$ for the era during which the gravitational waves are induced. In \texttt{SIGWfast.py} this value is hard-coded to $w = 1/3$ corresponding to radiation domination. In \texttt{SIGWfastEOS.py} this parameter can be specified through the variable \texttt{w}.
\item Set the flag \texttt{cs\_equal\_one}. In \texttt{SIGWfastEOS.py} the computation of $\OGW(k)$ can be done for a universe behaving like a perfect adiabatic fluid ($c_s^2=w$), or a universe whose energy is dominated by a canonical scalar field ($c_s^2=1$), see sec.~\ref{sec:output}. By setting \texttt{cs\_equal\_one = True} the computation is performed for the canonical scalar field case ($c_s^2=1$), while for the choice \texttt{cs\_equal\_one = False} it is the adiabatic perfect fluid ($c_s^2=w$) result that is computed. In the latter case the value of \texttt{w} in the previous step has to be chosen in the range $0< \,$\texttt{w}$\, <1$.
\end{enumerate}

\subsection{How to input the primordial scalar power spectrum $\Pzeta(k)$}
\label{sec:Pzeta}
These instructions apply to both \texttt{SIGWfast.py} and \texttt{SIGWfastEOS.py} and concern the block of code after `Configuration' and titled `Primordial scalar power spectrum'.

The principal input for SIGWfast is the primordial scalar power spectrum $\Pzeta(k)$. This can be provided as an analytical formula or in terms of numerical data:
\begin{itemize}
\item \textbf{Analytical formula} (\texttt{Num\_Pofk = False})\textbf{:} If an analytic scalar power spectrum is to be used as input, this is to be defined here as the function \texttt{Pofk(k)}. This should take a single argument which is the wavenumber $k$ and return the corresponding value of $\Pzeta$. Additional parameters of the power spectrum need to be declared as either global or local variables with given values. To keep the code as general as possible, there are no further restrictions on how \texttt{Pofk(k)} is to be defined. As long as calling the function \texttt{Pofk(k)} with a float argument returns a float, the script should run without any problems. The default example included with the code is the scalar power spectrum obtained for a strong sharp turn in the inflationary trajectory, see eq.~(2.25) in \cite{Fumagalli:2020nvq} and sec.~\ref{sec:examples} below.  

In SIGWfast, \texttt{Pofk(k)} is first discretized by evaluating it on an array of $k$-values, before an interpolation function is then used in the computation.\footnote{The reason is that SIGWfast uses `vectorization' to accelerate the computation. This means that a function, when provided with an array of values as input, outputs the corresponding result values also as an array. For vectorization to be applicable to \texttt{Pofk(k)} directly, certain rules in the definition of \texttt{Pofk(k)} would have to be followed. For example, piecewise-defined functions could not be defined via \texttt{if} statements, but constructions using e.g.~\texttt{numpy.where} would have to be used. Also, some functions that are provided in the \texttt{scipy} package are not vectorizable by default. The method of discretizing and then interpolating is thus used to not burden the user with such restrictions: \texttt{Pofk} can be defined without any vectorization requirements as the interpolation function is vectorizable in any case. Thus, for example, \texttt{if} statements can be used in the definition of \texttt{Pofk}, as can be seen in the default example provided.} The discretization is performed by evaluating \texttt{Pofk(k)} on $k$-values given in the array \texttt{kpzeta} which by default is an extended and denser version of \texttt{komega}.\footnote{\label{ftn:kpzeta}By default, \texttt{kpzeta} ranges from \texttt{kmin/2} to \texttt{2*kmax} and contains 4$\times$ as many entries as \texttt{komega}.} If needed, \texttt{kpzeta} can be defined here by the user in any other way.\footnote{Consider the following situation where the default definition of \texttt{kpzeta} would lead to wrong results. Take the case where $\Pzeta(k)$ is peaked around a value $k=k_\star$. For SIGWfast to produce meaningful results, \texttt{kpzeta} should span an interval that contains $k_\star$. At the same time one may be exclusively interested in the IR tail of $\OGW(k)$, hence choosing \texttt{komega} to only contain values $k \ll k_\star$. In this case \texttt{kpzeta} should be defined independently from \texttt{komega}.} For SIGWfast to produce meaningful results, \texttt{kpzeta} needs to be sufficiently dense so that the discretization of \texttt{Pofk(k)} faithfully captures the relevant features of $\Pzeta(k)$. 
\item \textbf{Numerical data} (\texttt{Num\_Pofk = True})\textbf{:} If numerical input is to be used for the scalar power spectrum, this is to be provided in a file \textquotesingle data/\textquotesingle +filenamePz+\textquotesingle .npz\textquotesingle. Here, `filenamePz' refers to any name chosen for this file by the user and which can be declared in step 4 of the configuration, see sec.~\ref{sec:config-step-by-step}. The~\texttt{.npz} file should be prepared to contain an array of $k$-values and an array of corresponding $\Pzeta$-values, which should be accessible via the keywords `karray' and `Pzeta', respectively. That is, after loading the data in this file via the command \texttt{Pdata  = numpy.load(\textquotesingle data/\textquotesingle +filenamePz+\textquotesingle .npz\textquotesingle)}, the arrays of $k$-values and $\Pzeta$-values should be given by \texttt{Pdata[\textquotesingle karray\textquotesingle]} and \texttt{Pdata[\textquotesingle Pzeta\textquotesingle]}, respectively.\footnote{To make the arrays of $k$-values and $\Pzeta$-values accessible via these keywords, the file needs to be prepared as \texttt{numpy.savez(\textquotesingle data/\textquotesingle +filenamePz, karray=kvalues, Pzeta=Pvalues)}, where \texttt{kvalues} and \texttt{Pvalues} refer to the arrays storing the corresponding values.} 
\end{itemize}
The script is now ready to be run!

\begin{figure}[t]
\centerline{
\includegraphics[width=0.62\textwidth,angle=0]{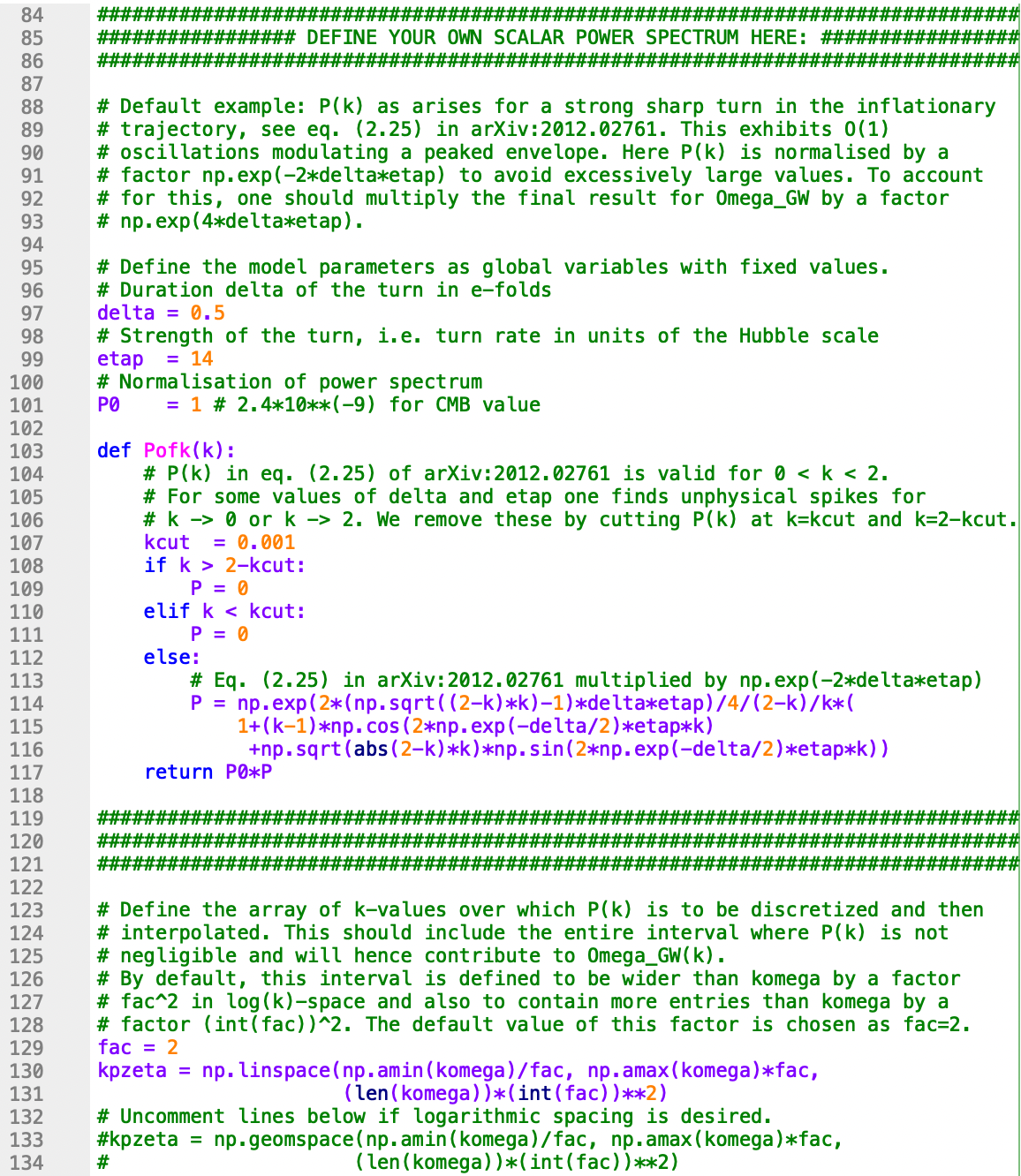}}
\caption{Section of code where the analytical expression for $\Pzeta(k)$ can be defined as the function \texttt{Pofk(k)}, here with the implementation of the formula given in eq.~\eqref{eq:Pzeta-strong-sharp-turn} as provided in the download versions of \texttt{SIGWfast.py} and \texttt{SIGWfastEOS.py}. Note the additional factor of $\exp(-2 \eta_\perp \delta)$ compared to \eqref{eq:Pzeta-strong-sharp-turn} to avoid excessively large values. In the lower part the default definition of the array \texttt{kpzeta} is shown, over which $\Pzeta(k)$ is discretized in the computation.}
\label{fig:Pofk-code}
\end{figure}

\section{Examples}
\label{sec:examples}
A great model for illustrating the functionality of SIGWfast is the contribution to the scalar power spectrum due to a strong sharp turn in the inflationary trajectory. This produces a localized peak in $\Pzeta(k)$ that exhibits further substructure in the form of oscillations. For a constant turn rate, the contribution to $\Pzeta(k)$ can be computed analytically and takes the form \cite{Palma:2020ejf,Fumagalli:2020nvq}:
\begin{align}
\label{eq:Pzeta-strong-sharp-turn}
\Pzeta(\kappa) = \ & \mathcal{P}_\textrm{env}(\kappa) \bigg[1 + (\kappa-1) \cos \Big(2 e^{-\frac{\delta}{2}} \eta_\perp \kappa \Big) + \sqrt{(2-\kappa)\kappa} \, \sin \Big(2 e^{-\frac{\delta}{2}} \eta_\perp \kappa \Big) \bigg] \times \Theta(2-\kappa) \, , \\
\nonumber & \textrm{with} \quad \mathcal{P}_\textrm{env}(\kappa) = \mathcal{P}_0 \frac{e^{2 \sqrt{(2-\kappa)\kappa} \, \eta_\perp \delta}}{4(2-\kappa) \kappa} \, , \quad \textrm{and} \quad \kappa = k / k_\star \, ,
\end{align}
where $\Theta(x)$ denotes the Heaviside theta function and $k_\star$ corresponds to the locus where the `envelope' $\mathcal{P}_\textrm{env}(k)$ of the scalar power spectrum peaks. The other parameters are \emph{(i)} $\delta$, the duration of the turn in $e$-folds, \emph{(ii)} $\eta_\perp$, the turn rate in Hubble units, and \emph{(iii)} $\mathcal{P}_0$, the `background value' of the power spectrum in absence of the turn.  

The formula \eqref{eq:Pzeta-strong-sharp-turn} is the default power spectrum implemented in the downloadable versions of both \texttt{SIGWfast.py} and \texttt{SIGWfastEOS.py}, see fig.~\ref{fig:Pofk-code} for the relevant lines of code. To avoid excessively large numbers due to the exponentially enhanced envelope, in the function \texttt{Pofk(k)} defined in the python scripts the envelope is normalised to unity at its maximum at $\kappa=1$. This is done by multiplying \eqref{eq:Pzeta-strong-sharp-turn} by an additional factor of $\exp(-2 \eta_\perp \delta)$.\footnote{This can be `undone' in the end by multiplying the computed result for $\OGW(k)$ by a factor $\exp(4 \eta_\perp \delta)$.} The argument of \texttt{Pofk(k)} corresponds to the variable $\kappa$ in \eqref{eq:Pzeta-strong-sharp-turn}, as we choose to measure $k$ in units of $k_\star$. The parameters $(\delta, \, \eta_\perp, \, \mathcal{P}_0)$ are implemented as the global variables (\texttt{delta}, \texttt{etap}, \texttt{P0}).

\begin{figure}[t]
\centerline{
\includegraphics[width=0.5\textwidth,angle=0]{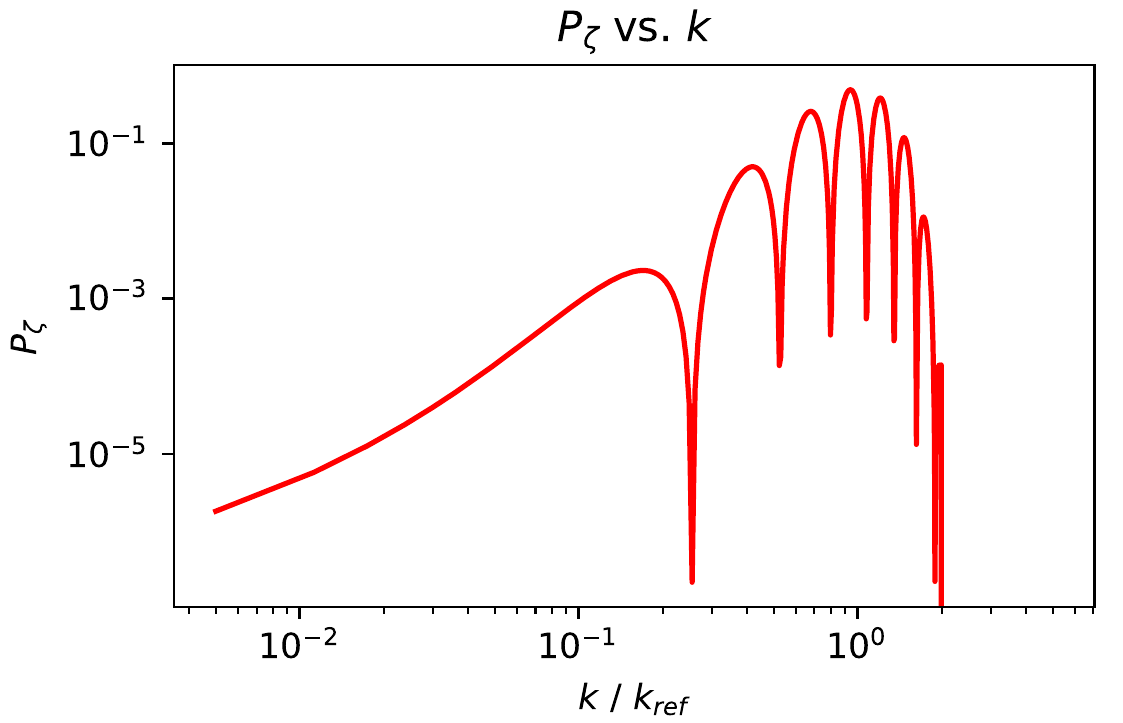}
\includegraphics[width=0.5\textwidth,angle=0]{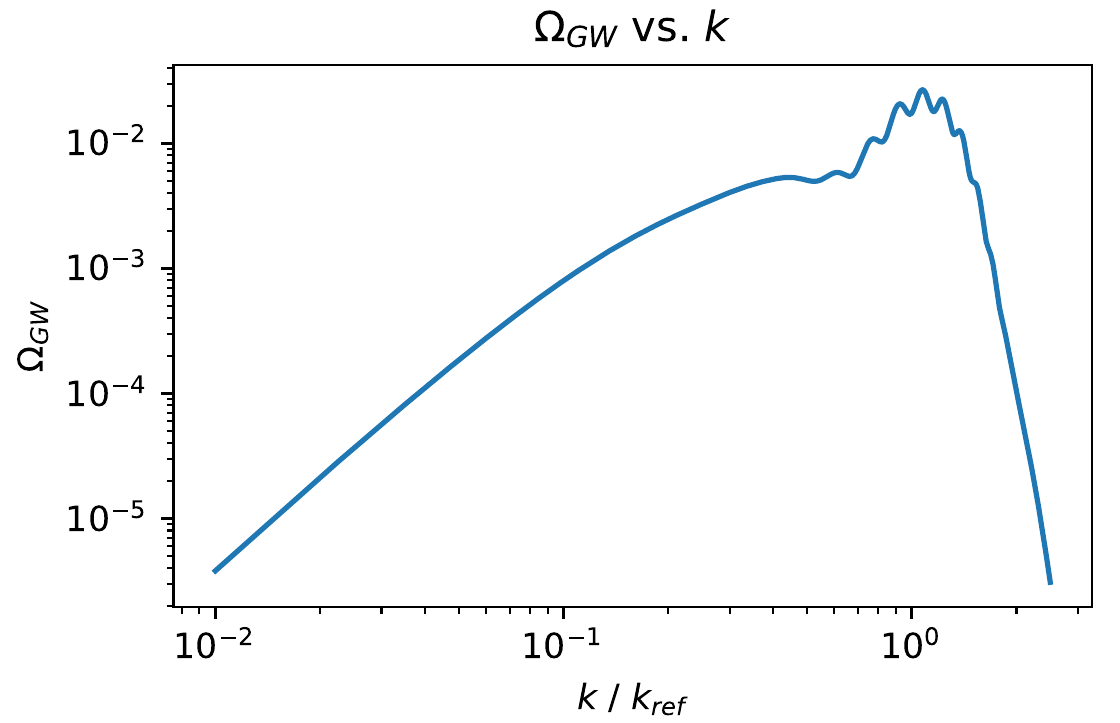}}
\caption{Scalar power spectrum $\Pzeta(k)$ (left panel) of the standard example in \eqref{eq:std-example} and the corresponding scalar-induced gravitational wave spectrum $\OGW(k)$ (right panel) as computed by \texttt{SIGWfast.py} (or \texttt{SIGWfastEOS.py} with \texttt{w = 1/3} and \texttt{cs\_equal\_one = False}). The unit $\kref$ corresponds to $k_\star$, the maxmimum of the envelope of $\Pzeta(k)$. Using the python-only version (\texttt{Use\_Cpp = False}) the computation takes 1.2s with \texttt{SIGWfast.py} and 2.8s with \texttt{SIGWfastEOS.py} on the development machine (Macbook Pro with M1 CPU). Using the compiled C++ module (\texttt{Use\_Cpp = True}) the same computation takes 0.9s with \texttt{SIGWfast.py} and 2.5s with \texttt{SIGWfastEOS.py}.}
\label{fig:P-and-OmegaGW-0p5-14-1}
\end{figure}

As the standard example we consider a moderately sharp turn with $\delta=0.5$ and choose $\eta_\perp=14$ to have an $\mathcal{O}(1)$ number of oscillations across the peak-region, see the left panel of fig.~\ref{fig:P-and-OmegaGW-0p5-14-1}. We further set $\mathcal{P}_0=1$ for simplicity, as its effect is just an overall rescaling:
\begin{empheq}[box=\graybox]{align}
\label{eq:std-example}
    \textrm{Standard example: \eqref{eq:Pzeta-strong-sharp-turn} with } \delta=0.5, \ \eta_\perp=14, \ \mathcal{P}_0=1, \textrm{ multiplied by a factor } e^{-2 \eta_\perp \delta} \, .
\end{empheq}

\begin{table}[t]
\begin{center}
\begin{tabular}{| l || c | c | c | c |}
  \hline			
  Number of points \texttt{nk} computed & 200 & 400 & 1000 & 2000 \\ \hline
  $t_{\textrm{comp}}$ of \texttt{SIGWfast.py} (python-only version) & 1.2s & 2.5s & 6.5s & 13.6s\\ \hline
  $t_{\textrm{comp}}$ of \texttt{SIGWfast.py} (with C++ module) & 0.9s & 1.9s & 5.0s & 10.7s \\ \hline
  $t_{\textrm{comp}}$ of \texttt{SIGWfastEOS.py} (python-only version) & 2.8s & 4.1s & 8.1s & 15.3s \\ \hline
  $t_{\textrm{comp}}$ of \texttt{SIGWfastEOS.py} (with C++ module) & 2.5s & 3.5s & 6.7s & 12.4s \\ \hline  
\end{tabular}
\end{center}
\caption{Comparison of computation times $t_{\textrm{comp}}$ of \texttt{SIGSWfast.py} and \texttt{SIGWfastEOS.py} (for the python-only version and the version using the compiled C++ module) for different values of \texttt{nk}, with the other configuration parameters as in fig.~\ref{fig:config-rad}. In \texttt{SIGWfastEOS.py} we have further set \texttt{w=1/3} and \texttt{cs\_equal\_one=False} to replicate the results of \texttt{SIGWfastEOS.py}. Computations were done on the development machine (Macbook Pro with M1 CPU) for the standard example \eqref{eq:std-example} as input for $\Pzeta(k)$. We observe that \texttt{SIGWfastEOS.py} is slower than \texttt{SIGWfast.py} by a constant offset of $\sim$1.6s, which is the time needed to load the more complicated integration kernel \eqref{eq:Twcs} that allows for different values of $w$. We also find that using the compiled C++ module generally  reduces computation times by up to $\sim 25\%$.}
\label{tab:comp-times}
\end{table}

\subsection{Gravitational waves induced during radiation domination}
\label{sec:examples-rad}
The simplest way of computing $\OGW(k)$ in this case is to use \texttt{SIGWfast.py}, which has been written for precisely this purpose. However, one can also use \texttt{SIGWfastEOS.py} as long as one also sets \texttt{w=1/3} and \texttt{cs\_equal\_one = False} in the configuration section.

We choose configuration settings as in fig.~\ref{fig:config-rad}. We have set \texttt{Num\_Pofk = False} to use the analytical formula for the scalar power spectrum. We wish to compute $\OGW(k)$ for wavenumbers in the range $k \in [0.01 k_\star, \, 2.5 k_\star]$, i.e.~setting \texttt{kmin = 0.01} and \texttt{kmax = 2.5}, which encloses the maximum of $\mathcal{P}_\textrm{env}$. We then define the array \texttt{komega} by covering this interval with \texttt{nk = 200} points that are linearly spaced (\texttt{numpy.linspace}), which we expect to provide a sufficient resolution of the result. The `normalization' factor $\mathcal{N}$ in \eqref{eq:OmegaGW-rad} or \eqref{eq:OmegaGW-w} has been set to unity (\texttt{norm = 1}) for simplicity.

When executing \texttt{SIGWfast.py} (or \texttt{SIGWfastEOS.py}), the script computes $\OGW(k)$ from the input scalar power spectrum and saves it in the file `data/OmegaGW\_of\_k.npz'. It then displays plots of both $\Pzeta(k)$ and $\OGW(k)$, which we show in fig.~\ref{fig:P-and-OmegaGW-0p5-14-1}. The following points are worth highlighting.
\begin{itemize}
\item 
The plot of $\Pzeta(k)$ in the left panel is an interpolation of the scalar power spectrum evaluated on \texttt{kpzeta}. Here we use the standard definition of \texttt{kpzeta}, see footnote \ref{ftn:kpzeta}. As this interpolation is used in the computation of $\OGW(k)$, it is important that it faithfully captures the relevant features of $\Pzeta(k)$, which one can check by inspecting the plot. Here we find that $\Pzeta(k)$ is indeed resolved sufficiently by the interpolation, including the oscillations, which are hence `visible' to the computation. The range of \texttt{kpzeta} is also wide enough to capture the entire region where $\Pzeta(k)$ is enhanced, which ensures that all these scales contribute to $\OGW(k)$. If this was not the case, we can either increase the range or resolution of \texttt{komega} (as this in turn sets \texttt{kpzeta} in the default case) or define a suitable \texttt{kpzeta} independently.  
\item The plot of $\OGW(k)$ in the right panel of fig.~\ref{fig:P-and-OmegaGW-0p5-14-1} is the main output of SIGWfast, and the corresponding data has been saved in the file `OmegaGW\_of\_k.npz' in the `data' subdirectory. The spectrum $\OGW(k)$ exhibits the characteristic morphology of gravitational waves induced by scalar fluctuations with a peak in $\Pzeta(k)$ at $k=k_\star$, i.e.~a broad bump around $k \sim 0.4 k_\star$ and principal peak around $k \simeq 2/ \sqrt{3} k_\star$. The most conspicuous property are however the oscillatory modulations of the principal peak, which are a direct consequence of the oscillations in $\Pzeta(k)$. See \cite{Fumagalli:2020nvq} for a detailed analysis of the origin and the properties of these modulations in $\OGW(k)$. The detection prospects of such an oscillatory gravitational wave spectrum with the upcoming observatory LISA has been studied in \cite{Fumagalli:2021dtd}.
\item Using the python-only version (\texttt{Use\_Cpp = False}) the computation takes 1.2s with \texttt{SIGWfast.py} and 2.8s with \texttt{SIGWfastEOS.py} on the development machine (Macbook Pro with M1 CPU). The same result can also be computed using the compiled C++ module by changing the corresponding flag in the configuration section to \texttt{Use\_Cpp = True}, which is available for machines running on MacOS or Linux. Executing the python script \texttt{SIGWfast.py} or \texttt{SIGWfastEOS.py}, a module named `sigwfast' is compiled, which takes $\mathcal{O}(1)$ seconds. The computation of $\OGW(k)$ itself is then shortened to 0.9s with \texttt{SIGWfast.py} and 2.5s with \texttt{SIGWfastEOS.py}. In any further runs with flag setting \texttt{Use\_Cpp = True} the compiled module is re-used and does not need to be recompiled, even if the input power spectrum or value of $w$ is changed. For a comparison of computation times as the number of points \texttt{nk} computed is changed, see table \ref{tab:comp-times}. 
\end{itemize}
We can also reproduce the results in fig.~\ref{fig:P-and-OmegaGW-0p5-14-1} using a numerical scalar power spectrum as input, by setting the flag \texttt{Num\_Pofk = True}, in which case the data for $\Pzeta(k)$ contained in the file `data/P\_of\_k.npz' is used. In the released version of SIGWfast this file contains data for the standard example \eqref{eq:std-example}.

\begin{figure}[t]
\centerline{
\includegraphics[width=0.5\textwidth,angle=0]{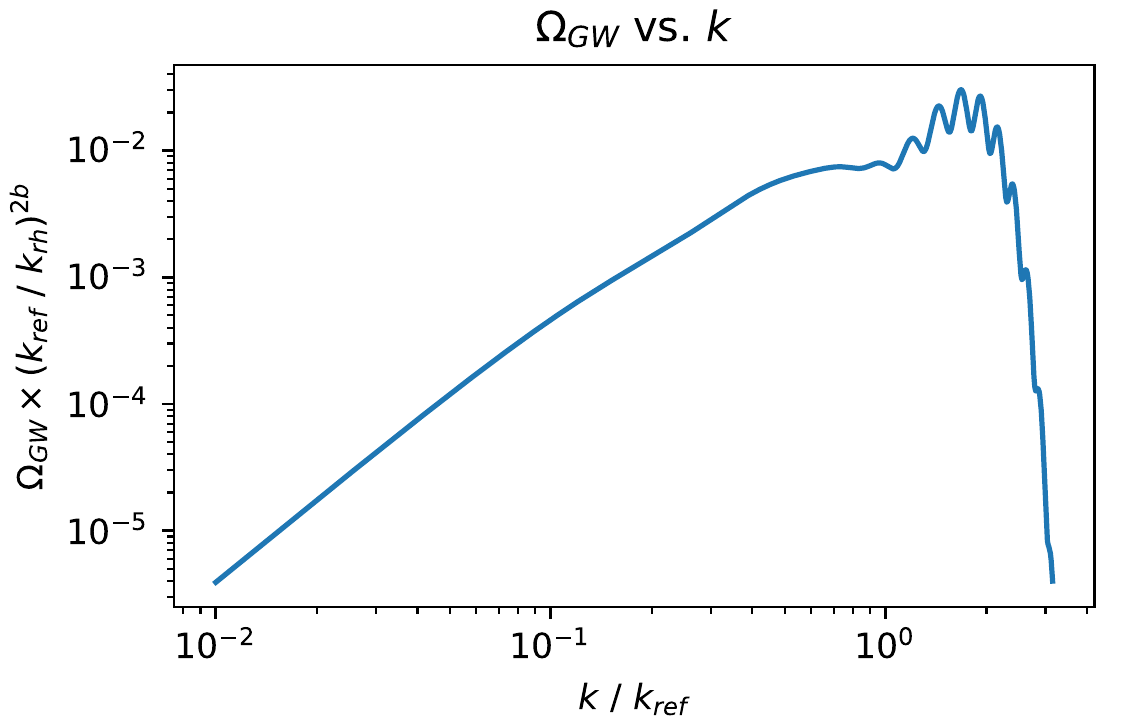}
\includegraphics[width=0.5\textwidth,angle=0]{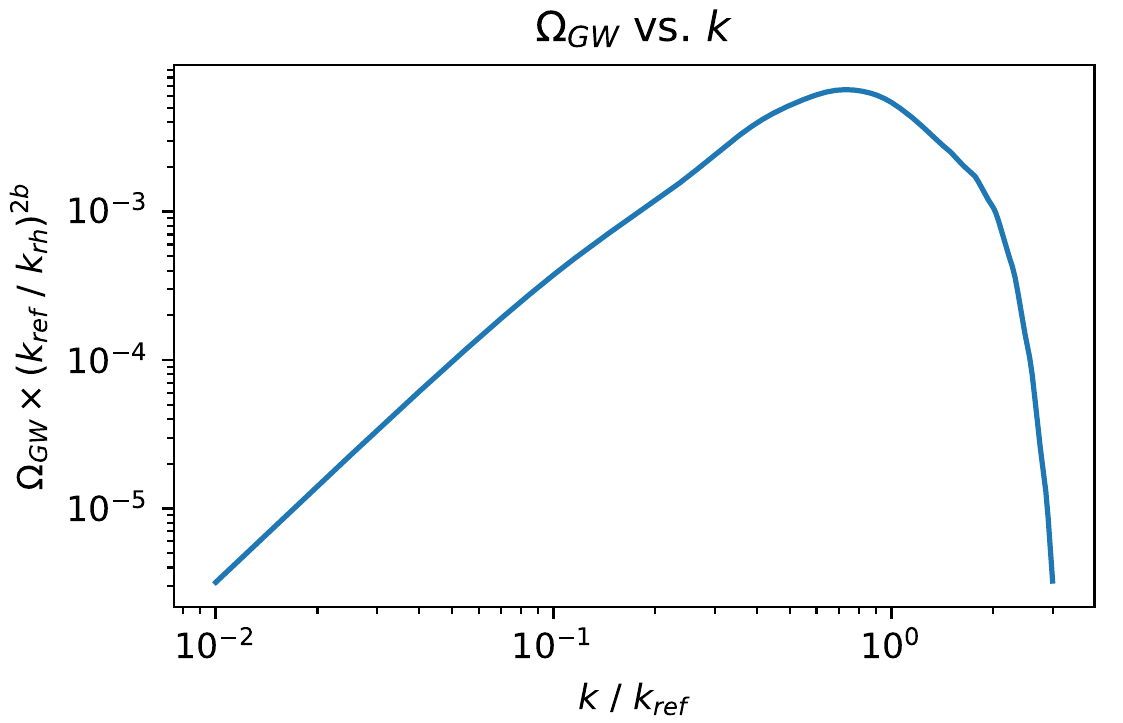}}
\caption{$\OGW(k)$ computed with \texttt{SIGWfastEOS.py} for \texttt{w = 0.8} with \texttt{cs\_equal\_one = False} (left panel) or \texttt{cs\_equal\_one = True} (right panel). The input scalar power spectrum is the same as shown in the left panel of fig.~\ref{fig:P-and-OmegaGW-0p5-14-1}. }
\label{fig:P-and-OmegaGW-0p5-14-1-0p8}
\end{figure}

\subsection{Gravitational waves induced during a phase with $w=0.8$}
\label{sec:examples-w}
Using the script \texttt{SIGWfastEOS.py} we can also explore the spectrum of gravitational waves induced in an era during which the universe evolved with an equation of state parameter $w$. We can adjust the equation of state parameter $w$ by giving a value to the variable \texttt{w} in the configuration section. By setting the flag \texttt{cs\_equal\_one} we can further consider a universe evolving as an adiabatic perfect fluid (\texttt{cs\_equal\_one = False}) or dominated by a canonical scalar field (\texttt{cs\_equal\_one = True}).

As input, we again consider the standard example for $\Pzeta(k)$ as given in \eqref{eq:std-example}. In fig.~\ref{fig:P-and-OmegaGW-0p5-14-1-0p8} we then show plots of $\OGW(k)$ produced by \texttt{SIGWfastEOS.py} for \texttt{w=0.8} and \texttt{cs\_equal\_one = False} (left panel) and \texttt{cs\_equal\_one = True} (right panel).\footnote{We slightly adjusted the limits of \texttt{komega} compared to those in the configuration in fig.~\ref{fig:config-rad} for better aesthetics.} On the development machine the computation times are $\sim$3s (left panel plot) and $\sim$1s (right panel plot), respectively. Recall that for $w \neq 1/3$ (i.e.~$b \neq 0$) the output of \texttt{SIGWfastEOS.py} is not $\OGW(k)$ today, but $\OGW(k) \times (\kref / \krh)^{2b}$, see sec.~\ref{sec:output-SIGWfastEOS}. 

The plots in fig.~\ref{fig:P-and-OmegaGW-0p5-14-1-0p8} show that the equation of state of the universe has an important effect on the scalar-induced gravitational wave spectrum. In the left panel one finds that in the adiabatic perfect fluid case the oscillations in $\OGW(k)$ for $w=0.8$ are even more pronounced than for radiation domination, cf.~the right panel of fig.~\ref{fig:P-and-OmegaGW-0p5-14-1}. Inspecting the right panel of fig.~\ref{fig:P-and-OmegaGW-0p5-14-1-0p8} one observes that in the canonical scalar field case $\OGW(k)$ does not exhibit any visible oscillations. For further reading on this topic we refer readers to \cite{Witkowski:2021raz}, where the effect of the expansion history of the universe on oscillatory signatures in $\OGW(k)$ has been analysed in detail. 

\section{Troubleshooting}
\label{sec:trouble}
SIGWfast has been written for python 3 and will not work with python 2. It has been developed using python 3.9.7 and `conda' for environment and package management, but has also been tested on python 3.8. The development machine was a Macbook Pro with a M1 CPU and running MacOS 12.1 Monterey. Python was installed in its x86 version and was running on the M1 chip via the Rosetta2 translator.\footnote{See the following \href{https://www.anaconda.com/blog/apple-silicon-transition}{blog post} on running python on Apple Silicon chips.} SIGWfast has also been tested on Ubuntu 20.04.4 and Windows Server 2019 running on a CPU with Intel x86 architecture.

\subsection{Compiling the C++ module}
\label{sec:trouble-Cpp}
One possible source of errors is the compilation of the C++ module. This is activated by setting the flag \texttt{Use\_Cpp = True} in the block of code titled `Configuration' and its use leads to a reduction in computation times of up to 25\%. This option is only available for systems running on Linux and MacOS. When trying to use the C++ option on Windows, the code automatically reverts to the python-only version.

On an older system running python 3.8 on MacOS 10.12 Sierra we encountered the problem that the automatic compilation of the C++ from the code was not initiated. As a result the module `sigwfast' could not be found and the computation ended with an error. To overcome this, the module `sigwfast' can be compiled by hand from the command line. It can then be used indefinitely, as it only has to be compiled only once. To do so, open the terminal and go to the `libraries' subfolder in the parent directory. For definiteness, here we assume that the parent directory `SIGWfast-main' is located in the home directory `$\sim$'. Hence, on the command line enter:
\begin{center}
\texttt{cd $\sim$/SIGWfast-main/libraries}
\end{center}
We have to work in this directory so that the file \texttt{SIGWfast.cpp} with the C++ code can be found. To compile the module by hand then enter:\footnote{If using python environments, ensure the correct environment is activated.}
\begin{center}
\texttt{python3 setup.py install --home=$\sim$/SIGWfast-main/libraries}
\end{center}
We used the command \texttt{python3} to make sure that python 3 is used, as the command \texttt{python} can sometimes refer to the version of python 2 that is shipped together with MacOS. The flag \texttt{--home=...} ensures that the module is deposited within the `libraries' subdirectory, rather than added to the other modules of the python distribution. This makes it easier to remove it later if desired. 


\section{Licensing}
\label{sec:license}
SIGWfast is distributed under the MIT license, a copy of which is included in the release. In case you have not received the license statement, see \href{https://spdx.org/licenses/MIT.html}{https://spdx.org/licenses/MIT.html}.

\section*{Acknowledgements}
We are indebted to Jacopo Fumagalli, without whom SIGWfast would have never been developed in this form and whose inputs vastly improved the code. We are also grateful to S\'{e}bastien Renaux-Petel, whose scientific insights and skilled leadership of the research group made the development of SIGWfast possible. Special thanks go to Guillem Dom\`{e}nech, whose research provided the scientific basis for SIGWfastEOS.py. We also thank John W. Ronayne, whose immense knowledge of python helped get this project off the ground. During the development of SIGWfast, L.T.W was supported by the European Research Council under the European Union's Horizon 2020 research and innovation programme (grant agreement No 758792, project GEODESI). 

\appendix
\section{Ferrers functions and associated Legendre function of the second kind}
\label{app:Legendre-definition}
Here we collect the definitions of $\mathsf{P}_{\nu}^{\mu}(x)$, $\mathsf{Q}_{\nu}^{\mu}(x)$ and $\mathcal{Q}_{\nu}^{\mu}(x)$ that appear in \eqref{eq:Twcs} as can be found in \cite{NIST:DLMF}: 
\begin{align}
    \mathsf{P}_{\nu}^{\mu}(x) &= {\bigg(\frac{1+x}{1-x} \bigg)}^{\mu/2} \frac{1}{\Gamma[1-\mu]} F \big(\nu +1, -\nu; 1-\mu ; \tfrac{1}{2}(1-x) \big) \, , \\
    \mathsf{Q}_{\nu}^{\mu}(x) &= \frac{\pi}{2 \sin(\pi \mu)} \Bigg[ \cos(\pi \mu) {\bigg(\frac{1+x}{1-x} \bigg)}^{\mu/2} \frac{1}{\Gamma[1-\mu]} F \big(\nu +1, -\nu; 1-\mu ; \tfrac{1}{2}(1-x) \big) \, , \\
    \nonumber & \hphantom{= \frac{\pi}{2 \sin(\pi \mu)} \Bigg[} - {\bigg(\frac{1-x}{1+x} \bigg)}^{\mu/2} \frac{\Gamma[\nu+\mu+1]}{\Gamma[\nu-\mu +1] \Gamma[1+\mu]} F \big(\nu +1, -\nu; 1+\mu ; \tfrac{1}{2}(1-x) \big) \Bigg] \, , \\
    \mathcal{Q}_{\nu}^{\mu}(x) &= \frac{\pi^{1/2} (x^2-1)^{\mu/2}}{2^{\nu+1}x^{\nu+\mu+1}} \frac{1}{\Gamma[\nu + 3/2]} F \big(\tfrac{1}{2} \nu + \tfrac{1}{2} \mu +1, \tfrac{1}{2} \nu + \tfrac{1}{2} \mu + \tfrac{1}{2}; \nu + \tfrac{3}{2} ; \tfrac{1}{x^2} \big) \, ,
\end{align}
where $F(a,b;c;x)$ is the Gauss hypergeometric function.

\bibliographystyle{JHEP}
\bibliography{Biblio}
\end{document}